# Observations of Fast Radio Bursts at Frequencies Down to 400 MHz

The CHIME/FRB Collaboration: M. Amiri[1], K. Bandura[2,3], M. Bhardwaj[4,5], P. Boubel[4,5], M. M. Boyce[6], P. J. Boyle[4,5], C. Brar[4,5], M. Burhanpurkar[7], P. Chawla[4,5], J. F. Cliche[4,5], D. Cubranic[1], M. Deng[1], N. Denman[8,9], M. Dobbs[4,5], M. Fandino[1], E. Fonseca[4,5], B. M. Gaensler[8,9], A. J. Gilbert[4,5], U. Giri[10,11], D. C. Good[1], M. Halpern[1], D. Hanna[4,5], A. S. Hill[1,12,13], G. Hinshaw[1], C. Höfer[1], A. Josephy[4,5], V. M. Kaspi[4,5], T. L. Landecker[12], D. A. Lang[10,11], K. W. Masui[14,15], R. Mckinven[8,9], J. Mena-Parra[4,5,14], M. Merryfield[4,5], N. Milutinovic[12,1], C. Moatti[4,5], A. Naidu[4,5], L. B. Newburgh[16], C. Ng[8], C. Patel[4,5], U. Pen[17], T. Pinsonneault-Marotte[1], Z. Pleunis[4,5], M. Rafiei-Ravandi[10], S. M. Ransom[18], A. Renard[8], P. Scholz[12], J. R. Shaw[1,12], S. R. Siegel[4,5], K. M. Smith[10], I. H. Stairs[1], S. P. Tendulkar[4,5,*], I. Tretyakov[19,8], K. Vanderlinde[9,8], P. Yadav[1]

---

* Corresponding Author
[1] Department of Physics and Astronomy, University of British Columbia, 6224 Agricultural Road, Vancouver, BC V6T 1Z1, Canada
[2] CSEE, West Virginia University, Morgantown, WV 26505, USA
[3] Center for Gravitational Waves and Cosmology, West Virginia University, Morgantown, WV 26505, USA
[4] Department of Physics, McGill University, 3600 rue University, Montréal, QC H3A 2T8, Canada
[5] McGill Space Institute, McGill University, 3550 rue University, Montréal, QC H3A 2A7, Canada
[6] Department of Physics and Astronomy, University of Manitoba, 301 Allen Building, 30A Sifton Road, Winnipeg, MB R3T 2N2, Canada
[7] Harvard University, Cambridge, MA 02138, USA
[8] Dunlap Institute for Astronomy and Astrophysics, University of Toronto, 50 St. George Street, Toronto, ON M5S 3H4, Canada
[9] Department of Astronomy and Astrophysics, University of Toronto, 50 St. George Street, Toronto, ON M5S 3H4, Canada
[10] Perimeter Institute for Theoretical Physics, 31 Caroline Street N, Waterloo, ON N2L 2Y5, Canada
[11] Department of Physics and Astronomy, University of Waterloo, Waterloo, ON N2L 3G1, Canada
[12] Dominion Radio Astrophysical Observatory, Herzberg Astronomy & Astrophysics Research Centre, National Research Council of Canada, P.O. Box 248, Penticton, BC V2A 6J9, Canada
[13] Space Science Institute, Boulder, CO 80301 USA
[14] MIT Kavli Institute for Astrophysics and Space Research, Massachusetts Institute of Technology, 77 Massachusetts Ave, Cambridge, MA 02139, USA
[15] Department of Physics, Massachusetts Institute of Technology, Cambridge, 77 Massachusetts Ave, MA 02139, USA
[16] Department of Physics, Yale University, New Haven, CT 06520, USA
[17] Canadian Institute for Theoretical Astrophysics, 60 St. George St, Toronto, ON M5S 3H8, Canada
[18] National Radio Astronomy Observatory, 520 Edgemont Rd., Charlottesville, VA 22903, USA
[19] Department of Physics, University of Toronto, 60 St. George St, Toronto, ON M5S 3H4, Canada

**Fast radio bursts (FRBs) are highly dispersed millisecond-duration radio flashes likely arriving from far outside the Milky Way galaxy[1,2]. This phenomenon was discovered at radio frequencies near 1.4 GHz and to date has been observed in one case[3] at as high as 8 GHz, but not below 700 MHz in spite of significant searches at low frequencies[4–7]. Here we report detections of FRBs at radio frequencies as low as 400 MHz, on the Canadian Hydrogen Intensity Mapping Experiment (CHIME) using the CHIME/FRB instrument[8]. We present 13 FRBs detected during a telescope pre-commissioning phase, when our sensitivity and field-of-view were not yet at design specifications. Emission in multiple events is seen down to 400 MHz, the lowest radio frequency to which we are sensitive. The FRBs show a variety of temporal scattering behaviours, with the majority significantly scattered, and some apparently unscattered to within measurement uncertainty even at our lowest frequencies. Of the 13 reported here, one event has the lowest dispersion measure yet reported, implying it is among the closest yet known, and another has shown multiple repeat bursts, as described in a companion paper[9]. Our low-scattering events suggest that efforts to detect FRBs at radio frequencies below 400 MHz will eventually be successful. The overall scattering properties of our sample suggest that FRBs as a class are preferentially located in environments that scatter radio waves more strongly than the diffuse interstellar medium (ISM) in the Milky Way.**

The CHIME telescope is located at the Dominion Radio Astrophysical Observatory (DRAO) near Penticton, BC, Canada and consists of four North-South (N-S) oriented, fixed 20 m × 100 m cylindrical reflectors each with a focal line populated with 256 equispaced antenna feeds digitized across 400–800 MHz. An on-site correlator[10,11] processes the 2,048 dual-polarization signals in real time, forming 1,024 independent intensity beams on the sky near transit. The input data from each beam consist of 16,384 frequency channels sampled at 1-ms cadence, which are searched by the CHIME/FRB instrument for dispersed, transient events. Only a small and variable number of beams were being searched during the pre-commissioning phase on which we report here. Full details on the CHIME/FRB system are provided elsewhere[8].

The data reported on here were obtained in July and August 2018, a period during which the CHIME/FRB instrument was not yet at final design specification. New racks of compute nodes were being installed and the number of nodes operating, and hence beams on sky, varied from day to day. Additionally, the number of compute nodes operating in the correlator, and hence the frequency coverage, also varied daily. Finally, calibration strategies, data pipeline configurations and data metric collection methods were also being tested.

In spite of the system's pre-commissioning status, we detected 13 FRBs during this period. Dynamic spectra for these events are shown in Figure 1. At least seven events are detected down to the lowest frequency of the CHIME band, some with no evidence for scatter-broadening within the time resolution of the instrument. However, eight of our sources show scattering. Detailed properties of these FRBs, measured as described in the Methods section, are shown in Table 1. Our events are broadband, with some spanning the entire 400 MHz of bandwidth, but others spanning a more limited range. Note that the observed spectrum is the convolution of the source's intrinsic spectrum with the highly frequency-dependent instrument response that has yet to be calibrated and which has structure on a wide variety of frequency scales. Moreover, the observed spectrum is also affected by the unknown location of the source in our primary beam (or possibly in a sidelobe). For these reasons, apart from asserting there is emission present where it is observed, currently we can say little more regarding the intrinsic source spectra, with one exception[9].

The burst widths reported in Table 1 were determined using our combination least-squares and Markov Chain Monte Carlo (MCMC) fitting routine (see Methods) and accounting for intra-channel dispersion smearing and scattering of the burst, both of which broaden the pulse with well-defined (but different) frequency dependencies. No attempt has been made to correct for bandpass non-uniformity or beam effects for this fit. We assume the pulse is a frequency-independent Gaussian convolved with a scattering-induced exponential tail, modeled at 1/16 of our instrumental resolution and boxcar-convolved. The widths are generally narrow compared to previously reported FRB widths. We attribute this to the well determined dispersion and scattering properties enabled by our large fractional observing bandwidth. This, and the assumed scattering index of –4, enable us to disentangle scattering, dispersion and intrinsic width, although in a model-dependent way. We caution that FRB pulses can show complex underlying structure[3,12], as discussed in our companion Letter[9].

That we have detected events having little to no apparent scattering throughout our band suggests that FRBs are detectable at radio frequencies well below 400 MHz, in spite of the lack of discoveries in this regime so far[4,5,7]. An event that is unscattered to within 1 ms at 400 MHz has at most 256 ms of scattering at 100 MHz, a regime which has been well-searched, but unsuccessfully, to date[13]. Although the redshift of each FRB determines how the emitted frequency compares to the observed frequency, the presence of 400-MHz emission in bursts with low excess dispersion measure (DM) and correspondingly low redshifts demonstrates that FRB environments can be optically thin to this emission, though a lower frequency cutoff is still possible. The event with the highest DM in our sample (1,007 pc cm$^{-3}$ for FRB 180817.J1533+42) also has the longest scattering time, and the three events with the lowest DMs have no measurable scattering. This hints at a possible DM/scattering correlation, however, there could be subtle selection biases in our detection pipeline. This is under study and will be discussed elsewhere. Little to no correlation is expected if the DMs of FRBs have a large contribution due to the intergalactic medium (IGM), where scattering is expected to be small[14,15]. We note that at least 6 FRBs are detected at the 800-MHz top of our band, while 8 are detected at the 400-MHz bottom of our band. Though our effective field-of-view is larger at the lower radio frequencies (an effect yet to be fully quantified), the prevalence of observable low-frequency emission in spite of the greatly enhanced scattering time does not support proposed low-frequency cutoffs due to free-free absorption or other physical mechanisms[16], at least within the CHIME band.

One of our events, FRB 180814.J0422+73, at DM 187 pc cm$^{-3}$ and high declination (+73º), shows repeat bursts, as detailed in an accompanying Letter[9]. It also has the widest burst of our sample, and some of its repeat bursts show complex structure[9].

Another of our events, FRB 180729.J1316+55, has a DM of only 109 pc cm$^{-3}$, the lowest yet observed for any FRB. The source was detected in three adjacent beams when our on-sky beam configuration had a single beam column in the N-S direction. We have made use of the 3-beam detection, together with studies of radio pulsar analogs, to improve the nominal localization of this source (see Methods), constraining the position to be (J2000) R.A. 13$^h$16$^m$ with uncertainty ±28', Dec. +55°36' with uncertainty ±8' (99%), where the R.A. uncertainty has been scaled by cos(Dec.) to reflect angular size. The maximum Galactic free electron column density in this direction is estimated[17] to be ~30 pc cm$^{-3}$, or[18] ~23 pc cm$^{-3}$, depending on the assumed model. Thus, the excess DM is in the range 80–90 pc cm$^{-3}$, though this may be further reduced due to the free electron content of the Galactic halo, which is estimated[19,20] to be ~30 pc cm$^{-3}$. Although the excess DM is not large, we believe this source is extragalactic as there is no evidence for any

Galactic sources of foreground DM such as HII regions[21] or star-formation regions[22]. Using an excess DM of 90 pc cm$^{-3}$ and models for the extragalactic free electron column as a function of redshift[23], the maximum possible redshift for FRB 180729.J1316+55 is 0.1, corresponding to a luminosity distance of 475 Mpc. This suggests that FRB 180729.J1316+55 is one of the closest known FRBs. The true redshift and distance are likely significantly smaller, as the host galaxy and environment local to the source probably contribute significantly to the source's DM.

We examined galaxies and radio sources in the localization box of FRB 180729.J1316+55 from the Sloan Digital Sky Survey, NRAO VLA Sky Survey, and Faint Images of the Radio Sky at Twenty-Centimeters (FIRST) survey. There are 78 galaxies with photometric redshifts $z < 0.1$ in the 99% confidence region. However, this number is limited by completeness of the surveys to low mass galaxies. The persistent radio source associated with the repeating FRB 121102 would have had a flux density of 0.9 mJy at a redshift of 0.1, detectable in the FIRST survey. We find 31 radio sources from the FIRST catalog (150 µJy rms) that lie in the 99% localization region. None of these can be confidently associated with FRB 180729.J1316+55 (see Methods).

Figure 2 shows scattering times referred to 1 GHz for our events compared to those of Galactic radio pulsars and previously detected FRBs. As previously noted[24,25], known FRBs are generally under-scattered relative to Galactic radio pulsars of the same DM, which likely reflects the low scattering in the intergalactic medium where it has been proposed that a significant portion of the observed FRB DM arises[14,15], as well as the fact that the signals from most Galactic pulsars must pass through the Milky Way disk to the reach the Earth. The CHIME/FRB events are even less scattered than the FRBs detected at higher frequencies, with one exception[26]. This is unsurprising given the steep power-law dependence of scattering time on radio frequency together with the bias our pipeline has against detection of highly scattered pulses, as demonstrated by our simulations (see Methods).

Even with this bias, it is clear that a majority of our events are scattered. We have performed an initial population synthesis analysis (see Methods) with a focus on scattering behaviour. The disks and spiral arms of large, Milky-Way-type galaxies are possible homes for FRBs if they originate from young stars, as is suggested by some models[27–29]. Our simulations (see Methods) of an FRB population in the disks of spiral galaxies having ISM and structure similar to that of the Milky Way show that this cannot reproduce the high observed scattering fraction in our CHIME/FRB events, even if the sources are located preferentially in a thin disk, the spiral arms, or concentrated within ~30 pc of the galactic centre. Our results therefore suggest that FRBs are preferentially located in environments with stronger scattering properties[30] than the quiescent diffuse ISM, although we cannot as yet distinguish between typical Galactic plane environments such as near HII regions and star-formation complexes, versus more extreme environments such as inside a very young supernova remnant or in the close vicinity of a supermassive black hole.

During pre-commissioning, we can only derive a 'floor' event rate in the CHIME frequency band based on the projected design sensitivity (though, as described above, the instrument was significantly less sensitive during this phase) and the number of beam-hours in the search. We refer to this as a floor because it is an approximate minimum for the observed all-sky event rate, but cannot be quantified with a statistically meaningful probability distribution. Our event rate floor (see Methods) is $3 \times 10^2$ events per day above a flux density threshold not lower than 1 (ms/$\Delta t$)$^{1/2}$ Jy, where $\Delta t$ is the pulse width including the intrinsic width, scattering, and intra-channel dispersion smearing. Our measured rate is comparable to the extrapolated lower limit ($4 \times 10^2$ events per day) from Connor et al. [31] and does not conflict with the upper limit of Chawla et al. [7].

The CHIME/FRB pre-commissioning phase was completed at the end of August 2018 with the deployment of the full 1,024 beams on sky. The system is currently in a commissioning mode with all 1,024 beams now processing data in real time. Full design capability of the instrument, including baseband data and public alerts, is anticipated in 2019.

**Acknowledgements** We are grateful for the warm reception and skillful help we have received from the Dominion Radio Astrophysical Observatory, operated by the National Research Council Canada. The CHIME/FRB Project is funded by a grant from the Canada Foundation for Innovation 2015 Innovation Fund (Project 33213), as well as by the Provinces of British Columbia and Québec, and by the Dunlap Institute for Astronomy and Astrophysics at the University of Toronto. Additional support was provided by the Canadian Institute for Advanced Research (CIFAR), McGill University and the McGill Space Institute via the Trottier Family Foundation, and the University of British Columbia. The Dunlap Institute is funded by an endowment established by the David Dunlap family and the University of Toronto. Research at Perimeter Institute is supported by the Government of Canada through Industry Canada and by the Province of Ontario through the Ministry of Research & Innovation. The National Radio Astronomy Observatory is a facility of the National Science Foundation operated under cooperative agreement by Associated Universities, Inc. P.C. is supported by an FRQNT Doctoral Research Award and a Mitacs Globalink Graduate Fellowship. M.D. acknowledges support from CIFAR, an NSERC Discovery and Accelerator Grants, and from FRQNT Centre de Recherche en Astrophysique du Québec (CRAQ). B.M.G. acknowledges the support of the Natural Sciences and Engineering Research Council of Canada (NSERC) through grant RGPIN-2015-05948, and the Canada Research Chairs program. A.S.H. is partly supported by the Dunlap Institute. V.M.K. holds the Lorne Trottier Chair in Astrophysics & Cosmology and a Canada Research Chair and receives support from an NSERC Discovery Grant and Herzberg Award, from an R. Howard Webster Foundation Fellowship from CIFAR, and CRAQ. C.M. is supported by a NSERC Undergraduate Research Award. J.M.-P. is supported by the MIT Kavli Fellowship in Astrophysics and a FRQNT postdoctoral research scholarship. M.M. is supported by a NSERC Canada Graduate Scholarship. Z.P. is supported by a Schulich Graduate Fellowship. S.M.R. is a CIFAR Senior Fellow and is supported by the NSF Physics Frontiers Center award 1430284. P.S. is supported by a DRAO Covington Fellowship from the National Research Council (NRC) Canada. FRB research at UBC is supported by an NSERC Discovery Grant and by CIFAR.

**Author Contributions** All authors on this paper played either leadership or significant supporting roles in one or more of: the management, development and construction of the CHIME telescope, the CHIME/FRB instrument and the CHIME/FRB software data pipeline, the commissioning and operations of the CHIME/FRB instrument, the data analysis and preparation of this manuscript.

**Competing Interests**  The authors declare that they have no competing financial interests.

**Correspondence**  Correspondence and requests for materials should be addressed to S. Tendulkar (email: shriharsh@physics.mcgill.ca).

| FRB | Width (ms) | DM (pc cm$^{-3}$) | DM$_{MW}$ (pc cm$^{-3}$) | R.A. (hh:mm) | Dec. (dd:mm) | Dec. FWHM (deg) | SNR | $\tau$ (ms) |
|---|---|---|---|---|---|---|---|---|
| 180725.J0613+67 | $0.31^{+0.08}_{-0.07}$ | $715.98^{+0.02}_{-0.01}$ | 71, 80 | 06:13 | +67:04 | 0.34 | 34.5 | $1.18^{+0.13}_{-0.12}$ |
| 180727.J1311+26 | $0.78 \pm 0.16$ | $642.07 \pm 0.03$ | 21, 20 | 13:11 | +26:26 | 0.35 | 14.2 | $0.6 \pm 0.2$ |
| 180729.J1316+55 | $0.12 \pm 0.01$ | $109.610 \pm 0.002$ | 31, 23 | 13:16 | +55:32 | … | 243.1 | $< 0.15$ |
| 180729.J0558+56 | $< 0.08$ | $317.37 \pm 0.01$ | 95, 120 | 05:58 | +56:30 | 0.32 | 25.2 | $< 0.26$ |
| 180730.J0353+87 | $0.42 \pm 0.04$ | $849.047 \pm 0.002$ | 57, 58 | 03:53 | +87:12 | 0.44 | 92.4 | $1.99 \pm 0.05$ |
| 180801.J2130+72 | $0.51 \pm 0.09$ | $656.20 \pm 0.03$ | 90, 108 | 21:30 | +72:43 | 0.35 | 41.1 | $5.0 \pm 0.3$ |
| 180806.J1515+75 | $< 0.69$ | $739.98 \pm 0.03$ | 41, 34 | 15:15 | +75:38 | 0.56 | 17.5 | $3.6 \pm 0.8$ |
| 180810.J0646+34 | $< 0.27$ | $414.95 \pm 0.02$ | 104, 140 | 06:46 | +34:52 | 0.33 | 17.7 | $< 0.40$ |
| 180810.J1159+83 | $0.28 \pm 0.03$ | $169.134 \pm 0.002$ | 47, 41 | 11:59 | +83:07 | 0.38 | 56.7 | $< 0.18$ |
| 180812.J0112+80 | $1.25^{+0.49}_{-0.47}$ | $802.57 \pm 0.04$ | 83, 100 | 01:12 | +80:47 | 0.38 | 19.8 | $1.9^{+0.3}_{-0.4}$ |
| 180814.J1554+74 | $< 0.18$ | $238.32 \pm 0.01$ | 41, 35 | 15:54 | +74:01 | 0.58 | 29.7 | $2.4 \pm 0.3$ |
| **180814.J0422+73** | $2.6 \pm 0.2$ | $189.38 \pm 0.09$ | 87, 100 | 04:22 | +73:44 | 0.35 | 24.0 | $< 0.40$ |
| 180817.J1533+42 | $< 0.37$ | $1006.840 \pm 0.002$ | 28, 25 | 15:33 | +42:12 | 0.32 | 69.9 | $8.7 \pm 0.2$ |

**Table 1: Properties of the pre-commissioning sample of CHIME/FRB events shown in Figure 1.** Names here are in format YYMMDD.Jhhmm+dd where YYMMDD is the UTC date of discovery and hhmm are beam R.A. hours and minutes, and +dd are degrees of Dec. Arrival times and fluence estimates for each event are reported in Tables 2 and 3, respectively, in Methods. R.A. and Dec. are the centres of the beams in which the event was detected. If an event was detected in more than one beam, we report data for the beam that had the highest SNR. Beam FWHM in R.A. is 19′ × secant(Dec.) at 600 MHz for all events. Beam FWHM in Dec. varies with zenith angle and is tabulated with reference to 600 MHz. Note that some sources were detected in their secondary transit, i.e. at zenith angle beyond the North Celestial Pole. We caution that the reported positions and beam FWHMs do not map directly to a confidence region for the location of the detected source, since the primary and synthesized beams have significant sidelobe structure which we are still working to characterize during the ongoing commissioning phase. Although in principle possible, with one exception, we do not report here on a refined localization based on detections in multiple beams. The position for FRB 180729.J1316+55, which was detected in 3 adjacent beams, was determined as described in the text and Methods. Its 99% uncertainty is ±21′ in R.A. after scaling by cos(Dec.) and ±8′ in Dec. An improved position for the repeater FRB 180814.J0422+73 (bold) is provided in our companion Letter[9]. DMs are measured assuming a dispersion index of –2. DM$_{MW}$ is the maximum DM due to the Milky Way; we provide values from the NE2001[17] and YMW16[18] models. SNR is the signal-to-noise ratio for the candidate derived from the MCMC fitting routine, which does not employ bandpass correction, nor does it account for beam responses. Hence SNR will not be linearly related to burst fluence (see Methods). $\tau$ is the scattering time at 600 MHz and is determined assuming a power-law frequency dependence (scattering index) of –4. All burst fits are done assuming a simple frequency-independent Gaussian burst profile. The values correspond to the median of MCMC samples with uncertainties enclosing the 68% credible interval. Upper limits are based on the 95% credible interval. For DM, an RMS systematic modelling error is added in quadrature to the statistical uncertainty.

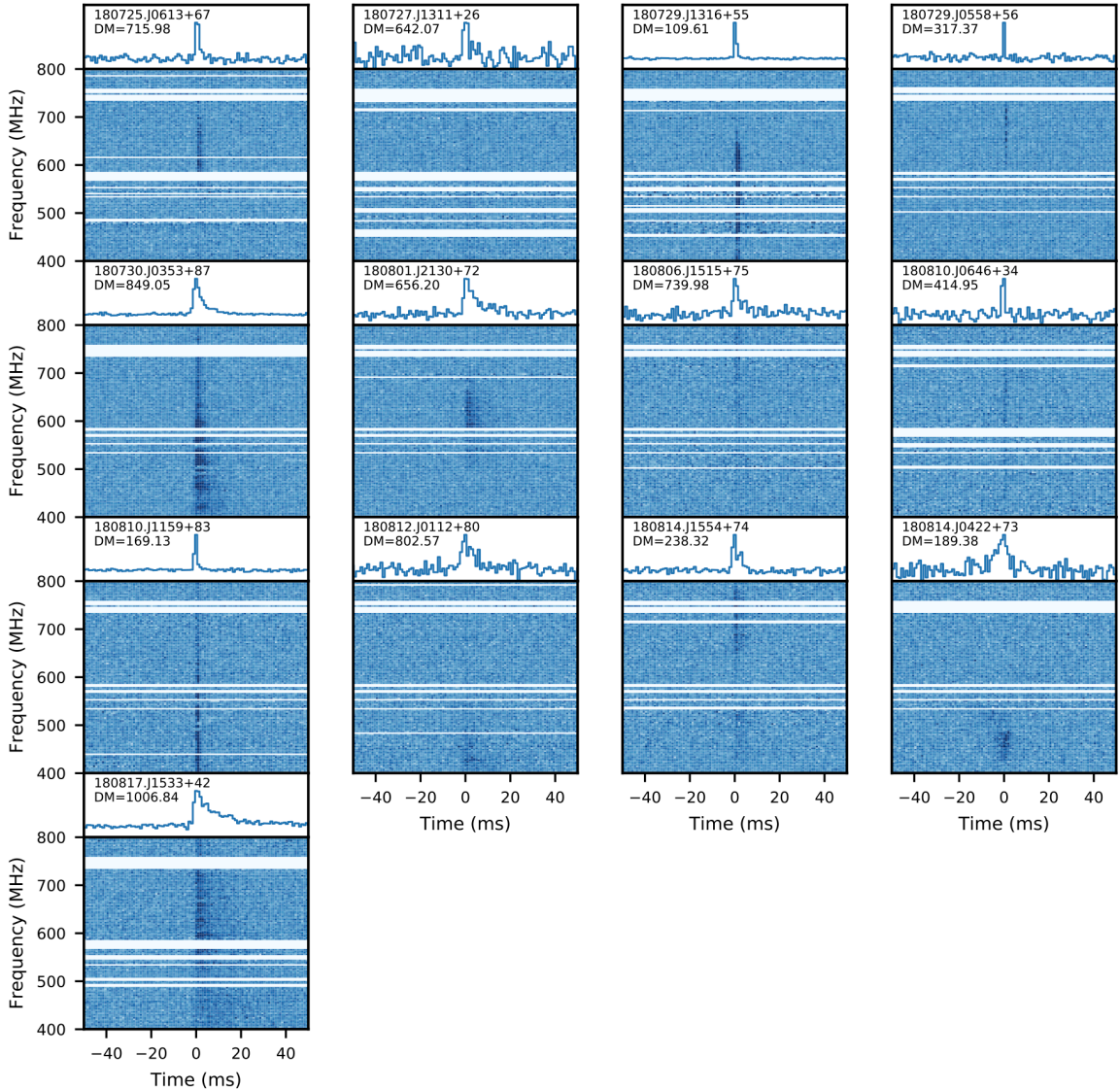

**Figure 1: Dynamic spectrum ("waterfall") plots for our sample of pre-commissioning CHIME/FRB events.** Colour-scale intensity is proportional to SNR, with all dynamic spectra using the same boundary values of intensity for colour mapping. The upper value of the boundary was chosen such that low-SNR signals are visible in their dynamic spectra, while saturating the spectra for high-SNR events. The blue line plots above each spectrum are frequency-summed burst profiles shown with full time resolution. Properties of individual events are provided in Table 1. For bursts detected in more than one beam, we show data from the beam with the highest SNR. We caution that the spectra shown here have not been calibrated for the effective bandpass, because this is strongly dependent on beam calibration — work in progress — and on the unknown location of the source in our primary beam or possibly in a sidelobe. Note that in nearly all cases radio frequencies in the 729–756 MHz band have been removed due to the presence of radio frequency interference from cell phone communication, as have narrower bands corresponding to television and other interfering signals (the bandwidth used for these plots ranges from 246 to 299 MHz).

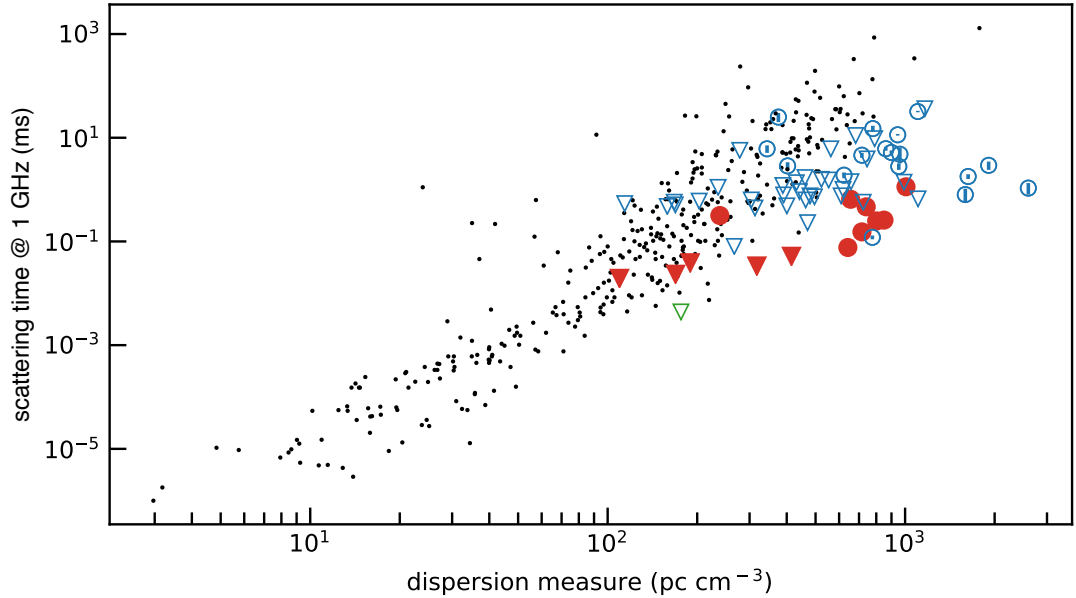

**Figure 2: Scattering time at 1 GHz versus DM.** Data points indicate CHIME/FRB events reported on here (red), other published FRBs (blue), and Galactic radio pulsars (black). CHIME/FRB scattering times have been extrapolated from the values in Table 1 assuming a scattering index of –4. Upper limits are represented by downward triangles, and error bars are present on FRB measurements but are generally smaller than the points. Pulsar data from the ATNF catalog (www.atnf.csiro.au/research/pulsar/psrcat/); FRB data from various sources as listed in the FRB catalog (http://frbcat.org) or in Ravi [32]. The green point is for FRB 170828[26], and is shown as an upper limit consistent with our requirement of detections to be at above the 99% confidence level.

# Methods

**CHIME Telescope Status:** As noted in the main article, the CHIME telescope was being commissioned during the period covered by this paper. There were several activities during this period that affected both the analog front-end of CHIME and the digital back-end, including repairing a handful of analog signal inputs and upgrading the F-engine firmware and X-engine software. In addition, the DRAO staff schedule occasional maintenance days during which time locally produced RFI might be expected to occur. Rather than detail all the telescope activities that occurred during the period covered by this paper, we annotate Extended Data Table 1 with a column of notes about activities that were taking place at the time each burst was detected.

One activity of note during this period was the initial deployment of a site survey radiometer for the Canadian Galactic Emission Mapper (CGEM) experiment: a 10 GHz survey project being developed at DRAO. On 180725, the radiometer was situated within the CHIME site and it was first powered on 20 minutes after the first CHIME FRB was detected (during a DRAO maintenance window). The team was concerned about the proximity of these events in time, but follow-up tests on the next day confirmed that this radiometer was not inducing any detectable RFI in the CHIME band. This equipment was permanently powered off after 26 July 2018.

**CHIME/FRB Phase Calibration and Beamforming System:** The CHIME correlator digitally forms 1,024 intensity beams, tiling the large (~250 deg$^2$) FoV of the CHIME telescope. Details of the beamforming scheme can be found elsewhere[33] but are summarized briefly below.

The 256 inputs corresponding to a single polarization along the focus of each cylinder are formed via Fast Fourier Transform into 256 beams distributed North-South (N-S), regularly spaced in sin $\theta$, where $\theta$ is the zenith angle.

These N-S beams are phased and summed across the four cylinders to form four East-West (E-W) sub-beams with the polarizations combined to measure intensity only. This results in a total of 1,024 beam pointings, which are static and do not track the moving sky. The polarisation-summed intensity data stream from each beam is subdivided into 16,384 frequency bins with a resolution of 24 kHz each, and downsampled to 0.983 ms. During this pre-commissioning stage, only one of the four N-S columns of beams was processed further.

The N-S (and, when available, E-W) beam locations are software configurable at run time. During pre-commissioning, they were configured to cover ±90° (i.e., horizon to horizon N-S).

**CHIME/FRB Calibration:** The FFT beamforming scheme employed in CHIME/FRB requires complex gain calibration[34] to be applied to the time stream from each feed prior to combining the signals into formed beams. During pre-commissioning, complex gain calibration was conducted using a simple scheme, using relative phase measurements between feeds derived from observations of Cygnus A on July 25, 31, Aug 3, and 12, 2018. During pre-commissioning, no attempt was made to correct for amplitude differences in feed response, time variations on scales faster than a few days, or the relative difference in phase due to feed-to-feed beam variations along different lines of sight.

To remove the phase contribution due to the analog system, we obtained a full set of $N^2$ visibilities for a transit of Cygnus A at the native 1,024 frequency-channel resolution used in the CHIME correlator. We eigen-decomposed the visibility matrix and extracted the eigenvectors corresponding to the two largest eigenvalues. We assume that the sky as seen in the $N^2$ visibilities is dominated by Cygnus A, and therefore the visibility matrix is rank 2, which allows us to interpret these eigenvectors as containing the complex gain for each input in the system, split by polarization[8,35]. This process is done for each frequency channel independently.

The phase contribution from interferometric spacing is described by a phase factor $e^{2\pi i \hat{\mathbf{n}} \cdot \mathbf{u}_{ij}}$, which depends only on the sky direction $\hat{\mathbf{n}}$ and the physical position of the antennas in u-v space, $\mathbf{u}_{ij}$ [31]. For our phase calibration, we calculate this factor relative to a single feed to generate a relative phase over the entire telescope. We then apply this factor to our vector of complex gains, fixing the amplitude of this vector at a constant value, and update it every few days. This vector is our final value for the complex gain of each antenna, which we apply prior to creating combinations of the signals in the FFT beamformer.

**CHIME/FRB Radio Frequency Interference Excision, Dedispersion and Initial Event Classification:** A detailed description of our real-time pipeline is provided elsewhere[8]. Briefly, the data stream from each beam is filtered for RFI-like signatures by applying a custom mask in frequency versus time space, using specialized RFI mitigation algorithms developed for CHIME/FRB[8]. The resultant data stream is incoherently dedispersed via a "tree" algorithm[36], and inspected for bursts. During pre-commissioning, we experimented with various RFI mitigation parameters and empirically converged on a configuration that had an acceptable false positive rate, as judged from the limited data then available. First, we downsampled each beam's data stream to 1,024 frequency channels and applied three iterations of nine clipping transforms (five standard-deviation transforms with $3\sigma$ threshold and four intensity transforms with $5\sigma$ threshold). The clipping sequence was followed by two detrending transforms: a degree 4 polynomial fit over time samples and an equivalent degree 12 spline over 1,024 frequency channels. The same set of twenty-seven clipping transforms was applied to the 1,024 intensity/weights once again before upsampling the mask to 16,384 frequency channels. Using the upsampled mask, the same set of detrending transforms was applied to the 16,384 intensity arrays prior to the dedispersion transform. The exact parameters of this RFI mitigation strategy are tunable and undergoing optimization during our commissioning phase.

In the relevant time period, our dedispersion and detection pipeline was operating using 6 dedispersion trees[8], which spanned a DM range of 0 to 13,000 pc cm$^{-3}$ and pulse widths up to 128 ms. Events having SNR>7 were reported in "coarse-grained"[8] samples having time and frequency resolution each 64 times the native resolution. During the relevant period, we searched in uniformly summed frequency channels (i.e. assuming 0 spectral index).

To mitigate spurious RFI signals that survive intensity cleaning and produce candidates, we employ two stages of machine learning classification (Support Vector Machines[37]). In both cases, the classifier predictions are expressed as a probability of astrophysical origin, which is interpreted as a grade. The first classifier operates on a per-beam basis and aims to capture SNR behaviour in the DM-time plane using two features. The first measures the SNR fall-off in the neighbourhood of a candidate's optimal parameters. The second feature counts the number of candidates with unique DMs in the previous ~10 sec. The second classifier operates after candidates from all beams have been collated and grouped in time, DM, and sky location. Each group is classified with the following features: the group's highest grade from the upstream classifier, the maximum SNR in the group, the ratio of the maximum SNR to the second highest value (if any), the total number of beams that harboured detections in the last ~10 sec, and the N-S extent of the group. Each classifier was trained with sets of hand-labelled candidate events — half identified as pulses from known pulsars, and the remaining as obvious RFI. In total, 10,000 and 40,000 events were used to train the first and second classifiers respectively. For validation, equally sized samples of labelled events were tested, resulting in 99% accuracy and recall. While these metrics are acceptable for the pre-commissioning phase, classification features are an active area of development.

**CHIME/FRB Event Identification:** A metadata pipeline ("L2/L3")[8] determines, in real-time, if events are likely astrophysical or RFI by grouping events that are close in time, DM and sky position and by considering multi-beam information. Properties of events that are deemed astrophysical are then compared with those of known pulsars, Rotating Radio Transients, and FRBs. Unknown sources are labelled "Galactic," "Ambiguous" or "Extragalactic" based on their DM and the maximum expected DM along their line-of-sight according to the NE2001[17] and YWM16[18] models of Galactic free electron density. We used a statistical criterion for classification in real time, in which a measured DM that exceeds the predicted Galactic values by $5\sigma$ is classified as extragalactic. (In this context, $\sigma$ is uncertainty and incorporates the systematic uncertainty of the Galactic models; for simplicity, we define systematic uncertainty to either be the difference between the two models or 25% of the measured value, whichever is largest.) A source with a DM that is greater than $2\sigma$ and less than $5\sigma$ is deemed "ambiguous;" and all sources with excess DM less than $2\sigma$ are classified as Galactic. Finally, post-processing actions are triggered based on a set of rules. At the epochs of the detections of the FRBs presented here, the pipeline was saving metadata (header) information for all detected events, including those labelled as RFI. Buffered full-resolution intensity data were saved to disk for beams in which an event was detected for all those associated with a known FRB or labelled "Extragalactic" or "Ambiguous" and which had SNR > 10.

**Localization Uncertainty of FRB 180729.J1316+55:** Since FRB 180729.J1316+55 was detected in 3 adjacent beams, the position can be improved using techniques similar to those presented[38] and applied[39] elsewhere. During pre-commissioning, there was only a single N-S column of beams active in the vicinity of the burst so improved localization using the multi-beam detection is possible only in declination. To constrain the burst's R.A., we selected 14 pulsars with declinations within 2° of the beam centre for the strongest detection. For the time span in which our reported FRBs were detected, 12,455 single pulses were recorded. By assigning the R.A. of these events according to the beam centre position at their time of arrival, we find 68% have values within 0.2° of their true R.A., and 99% within 0.8°. After scaling by cos(Dec.) to reflect angular extent on the sky, the 99% uncertainty is ±28'. For declination, we employ a frequency-dependent sensitivity model of our primary and formed beams. This model allows the ratios among recovered SNR values in adjacent beams to be compared against predicted ratios for a grid of positions and spectral indices. This comparison yields a narrow band of allowable declinations. When applying these methods to analogous detection patterns from pulsars, we find a subset that have large residuals which we interpret as systematic uncertainties of presently unknown origin. After quantifying these systematic uncertainties from these analogs, we constrain the 99% declination range to be ±12'.

**FRB Characterization:** Here we explain the provenance of the parameters in Table 1 for each reported FRB.

Using the time and DM from the real-time search as our initial guess, we start by simultaneously fitting a Gaussian pulse profile to the data using 16 frequency sub-bands across our 400-MHz bandwidth which quickly locates the burst and refines its time and DM and assigns an initial width. Next we use a frequency-independent Gaussian convolved with a frequency dependent one-side exponential as our model for the pulse. We fix the dispersion index to $-2$ and scattering index to $-4$, both of which are well motivated[40]. This was done to simplify fitting for this report; a more detailed analysis with these values fit for will be presented elsewhere. We compose the model pulse at an upsampled factor of 16 and 8 in time and frequency which we then boxcar-convolve to get the predicted pulse at our instrumental resolution. The likelihood function under this model is then optimized using least-squares fitting to get the best-fit model parameters. We further explore the parameter space around our best-fit values using MCMC sampling of our likelihood function. We use the Goodman-Weare[41] algorithm implemented in the public library emcee[42] to achieve this. After an initial burn-in phase, 500,000 samples are generated for each event. Our samples satisfy the commonly recommended convergence criterion based on integrated autocorrelation time and acceptance ratio. For low SNR events, we found that fitting in 4 sub-bands instead of 16 worked better.

We verified the accuracy of our fitting code using simulations that independently implement the same pulse model. In fitting to these simulations, we successfully recover the input parameters to within expectations based on statistical uncertainty. The exception is DM, which has an excess error of $0.0017$ pc cm$^{-3}$, which we attribute to differences in how the pulses are modelled below the time resolution. We add this error in quadrature to the statistical uncertainties. We also found that our ability to recover scattering, width, and DM are robust to changing the input spectrum, and hence issues such as the bandpass calibration do not bias our reported parameters.

The following describes specific individual source fits that are worthy of note.

**FRB 180727.J1311+26:** This event has signal in the bottom 3/4 of our band but low overall SNR, so its fit was done using only 4 frequency sub-bands.

**FRB 180729.J1316+55:** This event has the smallest DM of any currently known FRB and showed emission in the lower 3/4 of our band. Though there may be scattering, it is small and not well constrained. We report only an upper limit on the scattering time.

**FRB 180729.J0558+56:** This event shows emission in the upper half of the band, and is narrow. We report only upper limits on both width and scattering time.

**FRB 180810.J1159+83:** This event has a slightly asymmetric profile, but our fits do not demonstrate conclusively that there was scattering. We report an upper bound on scattering time.

**FRB 180812.J0112+80:** This event has most of its signal confined within the 400–500 MHz band. We fit the event using 4 sub-bands.

**FRB 180814.J0422+73:** This event has shown repeat bursts and is discussed in detail in an accompanying Letter[9].

**FRB 180817.J1533+42:** This event is observed throughout the band and has scattering time (at 600 MHz) $\tau > 8$ ms, the largest among the 13 events reported here.

**Estimation of Burst Fluences:** An accurate determination of burst fluence for the events reported here is complicated by the pre-commissioning status of the telescope wherein the system was changed faster than it could be characterized, by the non-traditional cylindrical design requiring the development of new calibration strategies, and because we have only begun the process of measuring the spatial and frequency response of our primary and synthesized beams. Beam measurements are significantly more challenging for a drift-scan telescope that cannot be scanned in elevation and azimuth across known calibration sources.

To simplify the fluence estimate, we assume all bursts were detected in the centre of the FFT-formed beams. For each event, we used bright sources with well known fluxes (such as Cygnus A, Cassiopiea A, Tau A, 3C 133, and others) located within 5° declination of the event for calibration. We assume North-South beam symmetry, so that sources on both sides of zenith can be used for each event. Since we did not record intensity data with the CHIME/FRB instrument for any of these point sources on the days the candidates were observed, we instead recorded data for the sources on a later date and for each frequency channel. We used this to obtain an estimate of the flux conversion as a function of frequency in the approximate direction of each candidate. The resulting calibration was applied to the spectrum of the FRBs and fluences were calculated (see Table 1), integrating over the usable bandwidth. We find that there are beam variations even within a 5° declination range of the event, leading to a significant error on the fluence measurements. In cases where we had multiple point sources near an event, we used these to assess fluence error by applying the calibration of one of the point sources to estimate the flux of the other. The typical error on the reconstructed flux is assessed as 44% and has been accounted for in the fluence measurement errors.

To account for the time variation of the system calibration, we derived flux calibrations for each of these sources over several days, and included this RMS variation in the reported fluence error. We also assessed the error associated with having used phase-only calibration, by reprocessing the point-source data with and without amplitude calibration, and included this in the error.

Extended Data Table 2 shows the fluence estimates and associated errors, as well as the bandwidth used therein. For these pre-commissioning events, typically about half of the total bandwidth is unavailable because some of the correlator processing nodes were offline or data transport from the F- to X-engine was corrupted, some of the frequency channels were contaminated with persistent RFI (about 20–25%), or the calibration solution failed for certain frequency channels. The fraction of usable bandwidth is expected to increase significantly during the commissioning phase.

**Simulations of Detection Pipeline:** To verify that the large fraction of our events showing scattering was not due to pipeline software detection bias, we used a pulse simulator to inject Gaussian pulses of width 1 ms but of various DMs and scattering times into our pipeline. First, for each value in a set of 8 logarithmically spaced scattering times ranging from 2 to 256 ms (referenced to 1 GHz), 100 simulated pulses were superimposed on white noise and injected. This was done for intrinsic SNRs 12 and 50 to simulate faint and bright pulses and for 3 DMs (600, 900 and 1,200 pc cm$^{-3}$). The average fraction of the pulse signal recovered by our detection code was compared for the various scattering times at fixed SNR and DM. As pulse scattering increased, less signal was recovered, with a >30% decrease for scattering times above 32 ms at 1 GHz, and over a factor of 2 loss above 128 ms. This injection analysis was repeated for pulses with no scattering using the same SNR values but with DMs that spanned the range thus far observed[43] (up to 2,500 pc cm$^{-3}$). We found less than 25% decrease in sensitivity in this range. Even out to DMs of 4,800 pc cm$^{-3}$ the drop was at most 30%. Thus, we conclude that the true fraction of CHIME/FRB events that are scattered is likely to be at least as high as the fraction we have measured for our pre-commissioning events.

**FRB Population Simulations:** We performed simulations to determine whether FRBs in the diffuse ISM of Milky Way-like host galaxies can reproduce the observation that 7 of our 13 events have scattering timescales at 600 MHz > 1 ms. In our simulations, we assume all FRBs originate from Milky Way type spirals, which we assume are viewed at random inclination angles. Prescriptions from the NE2001 model of Galactic electron density[17] are used to simulate the electron densities in the thick disk, thin disk, spiral arms and galactic center of the host galaxies. In each simulation run, we generate 13 events, integrating over the simulated electron density distribution from the event location to the near edge of the host galaxy. For a chosen galaxy distance, this allows estimation of dispersion measure, $DM_{host}$ and scattering measure, $SM_{host}$ for each event.

Initially, we assume all simulated events are located at 100 Mpc, approximately the distance estimated for the event with the lowest excess DM in our sample, FRB 180729.J1316+55. This assumption maximizes $SM_{host}$, as it is calculated by weighting the contribution of scattering material based on its location along the line of sight[17]. Additionally, we increase $SM_{host}$ by a factor of 3 (and by a factor of 6 in case of local scattering material) since plane waves from extragalactic sources exhibit greater pulse broadening per unit SM as compared to spherical waves from galactic sources[25,44]. We also account for the redshift, $z$, of the host galaxy by reducing $DM_{host}$ and $SM_{host}$ by $(1+z)$ and $(1+z)^3$, respectively[14,45]. To these, we add the Galactic DM and SM contribution by querying the NE2001 model for a random sky position sampled in a region which is centred on the corresponding CHIME/FRB event location and extends 2° in R.A. and 0.5° in Dec. We also add the IGM contribution to the DM by estimating it using the DM-redshift relation[23,45]. However, we do not simulate IGM scattering as it is estimated to be < 1 ms at 600 MHz, given a turbulence injection scale determined by AGN feedback[14,15]. We use the total simulated SM to compute a scattering timescale for each event under the assumption of a Kolmogorov spectrum and for a diffractive length scale smaller than the inner scale of turbulence[17,44].

We run the simulations until at least 50,000 runs have all 13 simulated scattering times < 128 ms (thus detectable with the CHIME/FRB pipeline) and simulated DM for each of the 13 events is less than the DM observed by CHIME/FRB for that event. Using statistics of runs that pass the selection criteria, we estimate the likelihood of obtaining scattering times at 600 MHz > 1 ms for at least 7 of the 13 events. We find at 95% confidence that a population of isolated FRBs cannot explain our observations, implying that FRBs must have a circumburst environment with strong scattering properties. We run the simulations with the local environment of FRBs having scattering properties similar to Galactic HII regions and supernova remnants[46] and find that the fraction of CHIME FRBs with scattering times > 1 ms at 600 MHz can be reproduced, provided these FRBs have a scale height similar to that of the pulsar population[47] (~300 pc) or are preferentially distributed along the spiral arms[48]. Although we do not model scattering properties of more extreme environments such as very young SNRs or regions close to the galactic center, we note that location of FRB progenitors in such environments could also be consistent with our observations.

We repeat the simulations for a distribution of redshifts for the host galaxies, determined by subtracting the simulated $DM_{host}$ and $DM_{MW}$ for each event from the observed DM of the corresponding CHIME/FRB event, assigning the remaining DM to the IGM and using the DM-redshift relation[23,45]. We find that the simulated FRB distribution has a large fraction of events at distances > 100 Mpc, thereby strengthening our conclusions since $SM_{host}$ is further reduced due to the line-of-sight weighting. We note that cool ionized circumgalactic clumps have been recently proposed as an alternative to strong scattering in the circumburst environment[49]. However, we do not attempt to simulate their contribution to scattering as it is strongly dependent on redshifts of the host galaxies and the composition of the circumgalactic media (CGM), both of which are not well constrained. Additionally, observation of scintillation in FRB 110523 constrained the scattering material to be within 44 kpc of the source[50], suggesting that the CGM might not be the dominant contributor to scattering of all known FRBs.

**Multi-wavelength Analysis of the FRB 180729.J1316+55 Field:** The field of FRB 180729.J1316+55 is in the coverage area for Sloan Digital Sky Survey (SDSS), NRAO VLA Sky Survey, and Faint Images of the Radio Sky at Twenty-Centimeters (FIRST) survey. We selected 6,026 objects identified as galaxies from the SDSS DR14 catalog. Based on $r$-band magnitude, every detected galaxy has a chance coincidence probability of > 50% of being in the 99% localization region of 0.37 deg$^2$. Of the 6,026 galaxies, 8% had photometric redshift fits that failed and 44% had photometric redshift errors > 0.1. The SDSS survey has a depth of $m_r$ < 22.7 mag, limiting the completeness of the survey to galaxies fainter than $M_r$ > −15.7 mag. Thus, the sample of 78 host galaxies with $z$ < 0.1 is expected to be significantly incomplete. Future deep imaging and multi-object spectroscopy of faint sources in this field is necessary.

The FIRST catalog has 31 detected sources that lie in the 99% confidence region. 17 of these have no associated counterparts in optical images. 8 of these FIRST sources are associated with SDSS galaxies. In particular one object, FIRST J131849.6+553227, is co-located with the bright center of a $M_R = -20.7$ mag galaxy at $z = 0.09$, but we find that the chance coincidence of such an object being in this localization region is high. The remaining 7 FIRST sources show faint optical emission in SDSS images upon visual inspection, but are not cataloged in the SDSS galaxy catalogs.

**Event Rate:** During pre-commissioning, the configuration of the telescope changed frequently and performance metric reporting was not fully implemented or debugged. Measurements of the instrument sensitivity and beams are not yet mature. This makes it impossible to quantify an event rate with statistically meaningful confidence intervals. Nevertheless, we can calculate a 'floor' on the event rate using the projected design sensitivity of the instrument, and the number of beam-hours over which the pre-commissioning search operated.

The approximate design sensitivity is estimated as follows. CHIME has 80 m × 80 m of instrumented collecting area and, approximating its aperture efficiency to be 50%, a formed beam thus has a forward gain of 1.16 K/Jy. Assuming the design system temperature of 50 K and 200 MHz of currently usable bandwidth, the $10\sigma$ detection threshold is 1.0 $(ms/\Delta t)^{1/2}$ Jy, where $\Delta t$ is the pulse width including the intrinsic width, scattering, and intra-channel dispersion smearing.

Several factors that have not been accounted for are expected to result in a significantly higher average detection threshold during the pre-commissioning period, including: the spatial and frequency response of the primary and formed beams, the reduction in primary beam response with increasing zenith angle, a system temperature that may be higher than the assumed value of 50 K, a lack of daily complex gain and amplitude calibration of the feed response, frequent changes in discarded bandwidth due to RFI and X-engine node up-time, and inefficiencies in our search pipeline.

We consider the sky rate at which we observed bursts above this threshold. Following Connor et al. [31], our 13 observed bursts imply a lower limit on the mean rate for repeat trials of identical surveys of 8.5 bursts at 95% confidence. An accounting of the fraction of FRB search nodes that were operational over the survey indicates that our exposure did not exceed 4600 beam-days, although it could have been substantially lower due to contamination by persistent sources of RFI and a smaller beam solid angle than the fiducial value of 0.256 deg$^2$. This number includes corrections for beams whose sky locations partially overlapped, an exclusion of observing time for which intensity data was not being saved to disk, and an exclusion of beams pointed more than 60º from zenith which were deemed to be too insensitive for inclusion. Using the assumed telescope collecting area, the beam solid angle for a single beam is 0.256 deg$^2$ at 600 MHz. We put a floor on the all-sky burst rate of $3 \times 10^2$ day$^{-1}$ for a flux density threshold of 1.0 $(ms/\Delta t)^{1/2}$ Jy.

**Data availability** The raw data used in this publication are available at https://chime-frb-open-data.github.io/.

**Code availability** The code used to characterize the discovered FRBs is available at https://chime-frb-open-data.github.io/.

| FRB | MJD (topocentric) | MJD (barycentric) | Notes |
|---|---|---|---|
| 180725.J0613+67 | 58324.74968534 | 58324.74686773 | a,b |
| 180727.J1311+26 | 58326.03616289 | 58326.03454033 | c |
| 180729.J1316+55 | 58328.03355599 | 58328.03178423 | - |
| 180729.J0558+56 | 58328.72798910 | 58328.72487751 | - |
| 180730.J0353+87 | 58329.15099464 | 58329.14985535 | d |
| 180801.J2130+72 | 58331.36614344 | 58331.36684263 | - |
| 180806.J1515+75 | 58336.59239707 | 58336.59163871 | - |
| 180810.J0646+34 | 58340.72840988 | 58340.72474620 | e |
| 180810.J1159+83 | 58340.94493626 | 58340.94368162 | e |
| 180812.J0112+80 | 58342.48996380 | 58342.48979586 | - |
| 180814.J1554+74 | 58344.59738935 | 58344.59692877 | - |
| **180814.J0422+73** | 58344.61791692 | 58344.61702927 | - |
| 180817.J1533+42 | 58347.07592826 | 58347.07557125 | - |

**Extended Data Table 1: Notes regarding CHIME site activity near epochs (which are referred to 600 MHz) of reported FRBs.** The arrival times reported for the repeater FRB 180814.J0422+73 are for its discovery observation. Additional arrival times for repeat bursts are reported in our companion Letter[9]. a - Maintenance window at DRAO, but no site maintenance activity reported. b - A prototype CGEM radiometer was powered on for the first time 20 minutes after this event was detected. See text for details. c - To test RFI, the CGEM radiometer was powered on 5 minutes prior to this event and was running until 42 minutes after. It was permanently powered off thereafter. d - Correlator GPU nodes were restarted 67 minutes prior to this event. System performance should be nominal. e - Several tests for human-induced RFI (from cell phones, key fobs, RF test equipment) took place on site for a few hours this day. This testing concluded 75 minutes prior to 180810.J0646+34.

| FRB | Fluence (Jy ms) | Integration Bandwidth (MHz) |
|---|---|---|
| 180725.J0613+67 | $12 \pm 7$ | 223 |
| 180727.J1311+26 | $14 \pm 10$ | 224 |
| 180729.J1316+55 | $34 \pm 18$ | 228 |
| 180729.J0558+56 | $9 \pm 5$ | 235 |
| 180730.J0353+87 | $50 \pm 32$ | 228 |
| 180801.J2130+72 | $28 \pm 20$ | 226 |
| 180806.J1515+75 | $24 \pm 17$ | 210 |
| 180810.J0646+34 | $11 \pm 7$ | 206 |
| 180810.J1159+83 | $17 \pm 11$ | 209 |
| 180812.J0112+80 | $18 \pm 12$ | 214 |
| 180814.J1554+74 | $25 \pm 17$ | 215 |
| **180814.J0422+73** | $21 \pm 15$ | 216 |
| 180817.J1533+42 | $26 \pm 15$ | 212 |

**Extended Data Table 2: Fluence estimates and associated errors of the pre-commissioning sample of CHIME/FRB events.** The reported uncertainties are derived from measurements of reconstructed flux variation with persistent point sources. They include the effects of the uncharacterized primary beam, the phase-only calibration, and time variation. The integration bandwidth is that used for the fluence calculation of each event. Frequency channels are excluded from the fluence calculation due to hardware issues, persistent RFI in some channels, or failure to obtain calibration solutions in some channels. The fraction of available bandwidth is expected to increase during commissioning.